\documentclass[fleqn,twoside]{article}
\usepackage{espcrc2}

\usepackage{graphicx}

\def\lab#1      {\hbox{\small #1} }
\newcommand{\be}{\begin{eqnarray}}
\newcommand{\ee}{\end{eqnarray}}
\newcommand{\ben}{\begin{eqnarray*}}
\newcommand{\een}{\end{eqnarray*}}
\newcommand{\la}{\langle}
\newcommand{\ra}{\rangle}

\def\mb#1         {\mbox{\boldmath $#1$}}
\def\diffn#1	  {\Delta^{-}_{#1}}

\newcommand{\AmS}{{\protect\the\textfont2
  A\kern-.1667em\lower.5ex\hbox{M}\kern-.125emS}}

\begin{document}

\title{Consistency of lattice definitions of $U(1)$ flux in Abelian 
projected $SU(2)$ gauge theory\thanks{Work supported in part by the 
U.S. Dept. of Energy under grant no. DE-FG05-01 ER 40617 
}}
\author{
Takayuki Matsuki$^{\lab{a} }$ \thanks{Presented by T. Matsuki.
Permanent address:
Tokyo Kasei University, 1-18-1 Kaga, Itabashi, Tokyo 173-8602, Japan}
 and Richard W. Haymaker
\address{
Dept. of Physics and Astronomy, 
Louisiana State University,
Baton Rouge, Louisiana, 70803 USA
}
}
\begin{abstract}
We reexamine the dual Abrikosov vortex under the requirement that the 
lattice averages of the fields satisfy exact Maxwell equations [ME]. The
electric ME accounts for the total flux and the magnetic ME determines the
shape of the confining string.  This leads to  unique and consistent
definitions of flux and electric and magnetic currents at finite lattice
spacing. The resulting modification  of the standard DeGrand-Toussaint
construction gives a magnetic current comprised of smeared monopoles.
\vspace{1pc}
\end{abstract}

\maketitle

In Abelian projected SU(2) lattice gauge theory in the maximal Abelian gauge
the existence of a dual Abrikosov vortex between quark and antiquark has
been well established\cite{sbh,bss,gips,koma}.   Here we tighten up the 
above picture further by incorporating the lattice Ward-Takahashi identity
derived from the residual $U(1)$ gauge symmetry\cite{dhh}. 
This is an Ehrenfest relation for the expectation 
value of the fields and currents giving  the electric Maxwell equations [ME]
exactly at finite lattice spacing.  This defines a unique lattice expression
for the field strength or flux.   

In the present work, we are  examining the impact of this on the study of the
dual Abrikosov vortex. The Ward-Takashi identity determines uniquely the
lattice definition of all components of the flux. To be consistent,
the magnetic ME must use the same definition. The standard
DeGrand-Toussaint\cite{dt} [DT] definition of the magnetic current is based
on a different definition of flux, resulting in inconsistencies in the
magnetic ME.

This consistency question is only relevant at finite lattice spacing and all
these concerns go away in the continuum limit. However the finite lattice
spacing effects  are significant for the values of $\beta$ that we often use
for calculations.

Let us consider three definitions of field strength, all agreeing to lowest
order in $a$. The first definition was used by DT to define monopoles:
\ben
 \widehat{F}_{\mu \nu}^{(1)}&=&\theta_\mu(\mb{m} )-\theta_\mu(\mb{m} +\nu)\\
  &&
  - \theta_\nu(\mb{m} ) + \theta_\nu(\mb{m} +\mu) - 2 \pi n_{\mu\nu},\\
  &\equiv& \theta_{\mu \nu}(\mb{m} ) - 2 \pi n_{\mu \nu},\\
  &=& \overline{\theta}_{\mu \nu}(\mb{m} ),
\een
where $\theta_\mu(\mb{m} )$ refers to the $U(1)$ link angle in the domain
$-\pi < \theta_\mu < +\pi$.  The integers $n_{\mu \nu}$
are determined by requiring that  $-\pi <\overline{\theta}_{\mu \nu} < +\pi$.
($\overline{\theta}_{\mu \nu}$ is a periodic function of $\theta_{\mu \nu}$
with period $2 \pi$, i.e. it is a ``sawtooth function".)

The second and third  gives the exact electric ME for lattice averages
\be
\Delta_\mu^{-}
\la  \widehat{F}_{\mu\nu}^{(i)}\ra_W &=&
 \la \widehat{J}_\mu^{e(i)} \ra_W, \quad \quad i = 2,3,
\label{judy}
\ee
where 
\ben
\la \cdots \ra_W = \frac{\la \cdots e^{i \theta_W}\ra}{\la e^{i \theta_W}\ra}.
\een
For a pure $U(1)$ theory with a Wilson action, and for an Abelian
projected SU(2) theory with Wilson action respectively :
\be
  \widehat{F}_{\mu \nu}^{(2)} &=& \sin \theta_{\mu \nu}. 
\label{ben}\\
  \widehat{F}_{\mu \nu}^{(3)} &=& C_\mu(\mb{m} )C_\nu(\mb{m}
  +\mu)C_\mu(\mb{m} + \nu)
  C_\nu(\mb{m} )\times \nonumber \\&& \sin \theta_{\mu \nu}. \label{eric}
\ee
where the  $C$  factors are associated with the matter field  in
the Abelian projection, as explained below.  Here quantities
with $~\widehat{ }~$ mean those which appear in the lattice numerical
calculation without appending factors of $e$ and $a$.  

To derive Eqn.(\ref{judy}) for the $U(1)$ case consider
\ben
Z_W(\epsilon_\mu(\mb{m} )) = \int [d \theta] \sin \theta_W
\exp\left( \beta S
\right).
\een
This is invariant under the shift of any link angle
\ben
\theta_\mu (\mb{m} ) \rightarrow \theta_\mu(\mb{m} ) +  \epsilon_\mu(\mb{m} )
\een
which leads to Eqn.(\ref{judy})\cite{zfks}.  For the $SU(2)$ case\cite{dhh} we
use instead the fact that the Haar measure is invariant under 
multiplication by an infinitesimal group element $1 + i \epsilon_3 \sigma_3$.
The definition of flux that emerges involves the diagonal elements of the
$SU(2)$ links
\ben
\widehat{F}^{(3)}_{\mu \nu} &=&  
\frac{1}{2i}Tr(\sigma_3 
D_\mu(\mb{m} ) 
D_\nu(\mb{m} + \mu) \times\\&&
D_\mu^\dagger(\mb{m} + \nu) 
D_\nu^\dagger(\mb{m} ) ),
\een
where
\ben
D_{\mu}(\mb{n} ) 
&=& 
\left( 
\begin{array}{cc}
C_{\mu}  e^{i \theta_{\mu} }  
& 0\\
0&
C_{\mu}  e^{-i \theta_{\mu} }
\end{array}
\right).
\een
Working in a fixed gauge significantly complicates the derivation since the
shift of a link takes one out of the gauge. A compensating infinitesimal gauge
transformation restores the gauge.  

Having defined a unique flux $\widehat{F}_{\mu \nu}^{(i)}$ through the
electric ME, the magnetic ME is
\ben
 \frac{1}{2}\epsilon_{\mu\nu\rho\sigma}\Delta_\nu^{+}
  \widehat{F}_{\rho\sigma}^{(i)} &=&
  \widehat{J}_\mu^{m(i)} \quad \quad i = 2,3
\een
However the standard DT definition of current is 
\ben
  \widehat{J}_\mu^{m(1)} &=&
 \frac{1}{2}\epsilon_{\mu\nu\rho\sigma}\Delta_\nu^{+}
  \widehat{F}_{\rho\sigma}^{(1)} 
\een
and hence if we use the conventional  $\widehat{F}^{(1)}$ to define the
monopole current, and $\widehat{F}^{(2)}$ or $\widehat{F}^{(3)}$ respectively
for $U(1)$ and $SU(2)$ theories to get an exact expression for flux in the
confining string, then the magnetic ME is violated.  

The solution is to relax the requirement that we use the DT monopole
definition and use $\widehat{F}^{(2)}$ or $\widehat{F}^{(3)}$ instead when
defining monopoles. A simple configuration for the $U(1)$ case
($\widehat{F}^{(2)}$) will illustrate the effect.  Consider a single 
DT monopole with equal flux out of the six faces of the cube.  Then the ratio
of the $\widehat{F}^{(2)}$ flux out of this
cube compared to the DT $\widehat{F}^{(1)}$ flux gives
\ben
\frac{6 \sin (2 \pi/6)}{6 (2 \pi/6)} \approx 0.83.
\een
Since the current is conserved, the balance is made up by magnetic charge in
the neighboring cells. On a large surface the total flux is the same for the
two definitions.

The electric ME determines the total flux in the confining string and the
magnetic ME determines the transverse shape.  Further the latter enters
directly in the determination of the London penetration length, $\lambda_d$.
To see this let us consider the classical Higgs theory which we use to model
the simulation data.  The dual field is given by
\ben
\widehat{G}_{\mu \nu}(\mb{m} )  &=&
 \Delta_\mu^{-} \theta_\nu^{(d)}(\mb{m} ) -
 \Delta_\nu^{-} \theta_\mu^{(d)}(\mb{m} ),
\een
where $\theta_\mu^{(d)}(\mb{m} )$ is a dual link variable.  
Let us choose to break the gauge symmetry spontaneously 
through  a constrained Higgs field.
\ben
  \Phi(\mb{m} ) = v \rho(\mb{m} ) e^{i\chi(\mb{m} )},
  \quad \quad \rho(\mb{m} )=1.
\een
Under these conditions the magnetic current simplifies to
\ben
  \widehat{J}^m_\mu (\mb{m} ) = 2e_m^2 v^2
  \sin\left\{
  \theta_\mu(\mb{m} )+ 
  \chi_\mu(\mb{m} + \mu) -\chi_\mu(\mb{m} ) 
  \right\}.  
\een
where $e_m$ is the magnetic coupling. The magnetic ME is
\be
  \Delta_\mu^+ \widehat{G}_{\mu \nu} &=& \widehat{J}_\nu^m .
\label{fred}
\ee

For small $\theta^{(d)}$ it is easy to see that there is a London relation of
the form
\be
  \widehat{G}_{\mu \nu}(\mb{m} ) 
  = \frac{1}{2e_m^2 v^2}\left(\Delta^-_\mu \widehat{J}^m_\nu (\mb{m} ) - 
  \Delta^-_\nu \widehat{J}^m_\mu (\mb{m} )\right).
\label{bob}
\ee
Taking the confining string along the $3$ axis and choosing $\mu = 1$ and
$\nu = 2$ we see that the profile of the third component of curl of the
magnetic current must match the third component of the electric flux profile.
This assumes an infinite Higgs mass $M_H$. With a finite mass there is a
transition region of size $\sim 1/M_H$ in the core of the vortex but the
above London relation must hold sufficiently far outside the core.

Combining Eqns.(\ref{fred}) and (\ref{bob}) we get the relation
\ben
  \left(1-\lambda_d^2
\Delta_\mu^+ \Delta_\mu^- \right) \left<\widehat{E}_3(\mb{m} )\right>_W =0,
\lambda_d^{-1} =  e_m v\sqrt{2}
\label{mary}
\een
{\em The corresponding equations in the simulation of the 
$SU(2)$ theory must also be satisfied in order to arrive 
at this correct expression for $\lambda_d$ and hence the importance of our 
definitions.}  

We generated $208$ configurations  on a $32^4$ lattice at $\beta = 2.5115$.
The maximal Abelian gauge fixing used over-relaxation.   Fig.(\ref{mars})
shows the profile of the electric flux corresponding to $\widehat{F}^{(2)}$
and $\widehat{F}^{(3)}$. Fig.(\ref{venus}) shows the profile of the theta
component of the magnetic current corresponding to $\widehat{F}^{(1)}$,
$\widehat{F}^{(2)}$ and $\widehat{F}^{(3)}$.

In summary, we showed that consistency requires one use the same definition of
flux throughout.  If, for example, one uses  $\widehat{F}^{(3)}$ definition of
electric field (bottom graph in Fig.(1)) in order to account correctly for the
total flux but then uses the DT definition of current (top graph in Fig.(2))
we would then incur errors of $\sim 40$\%.
%
\label{mars}
\includegraphics[trim=0 20mm 0 20mm,scale=0.35]{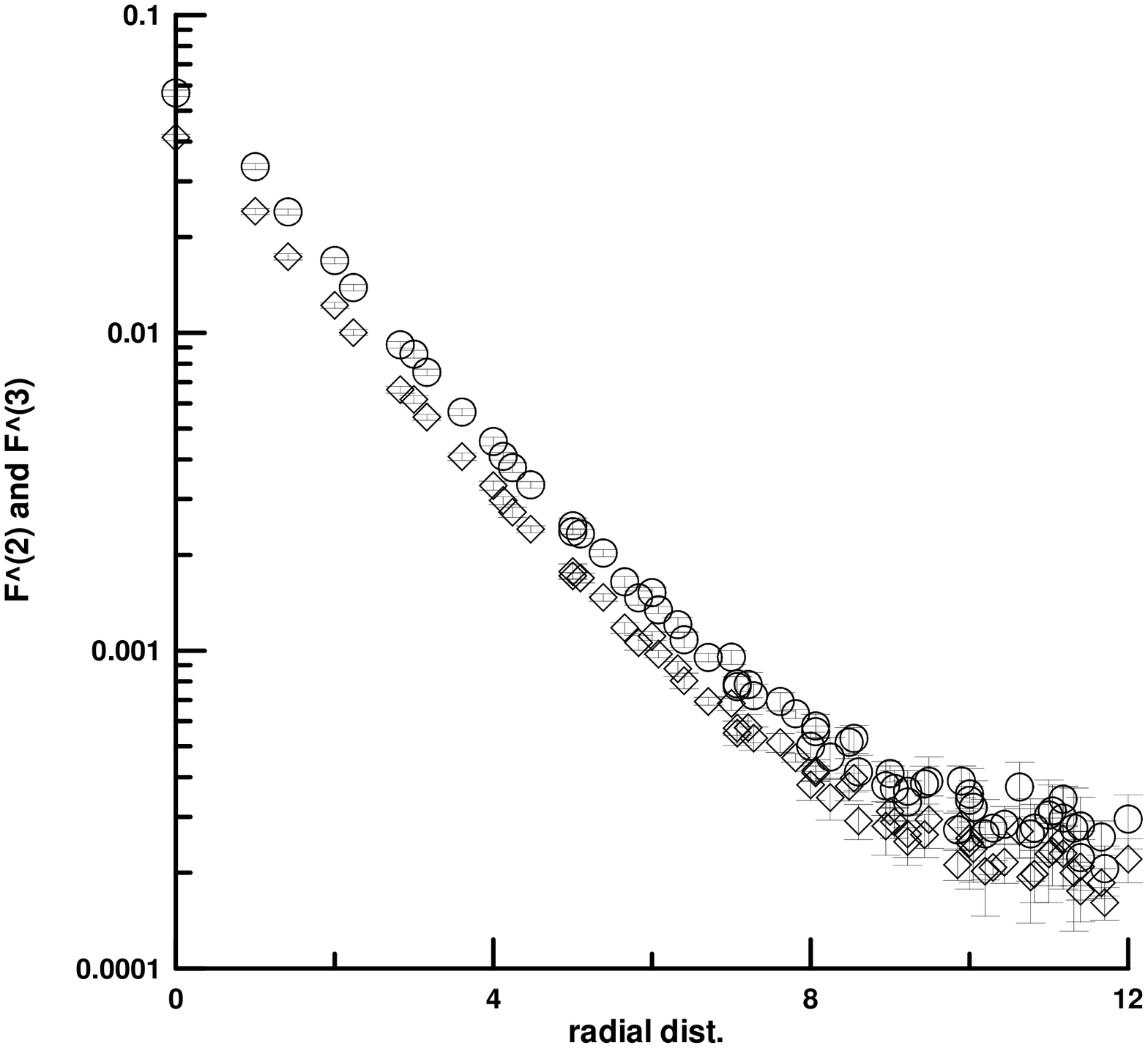}
Figure 1. Profile of the electric field (highest to lowest)
$\widehat{E}_3^{(2)}$ (circle) and $\widehat{E}_3^{(3)}$ (diamond)
on the mid-plane 
between $q$ and $\bar{q}$ separated by $13a$.
%
\label{venus}
\includegraphics[trim=20 35mm 20 30mm,scale=0.40]{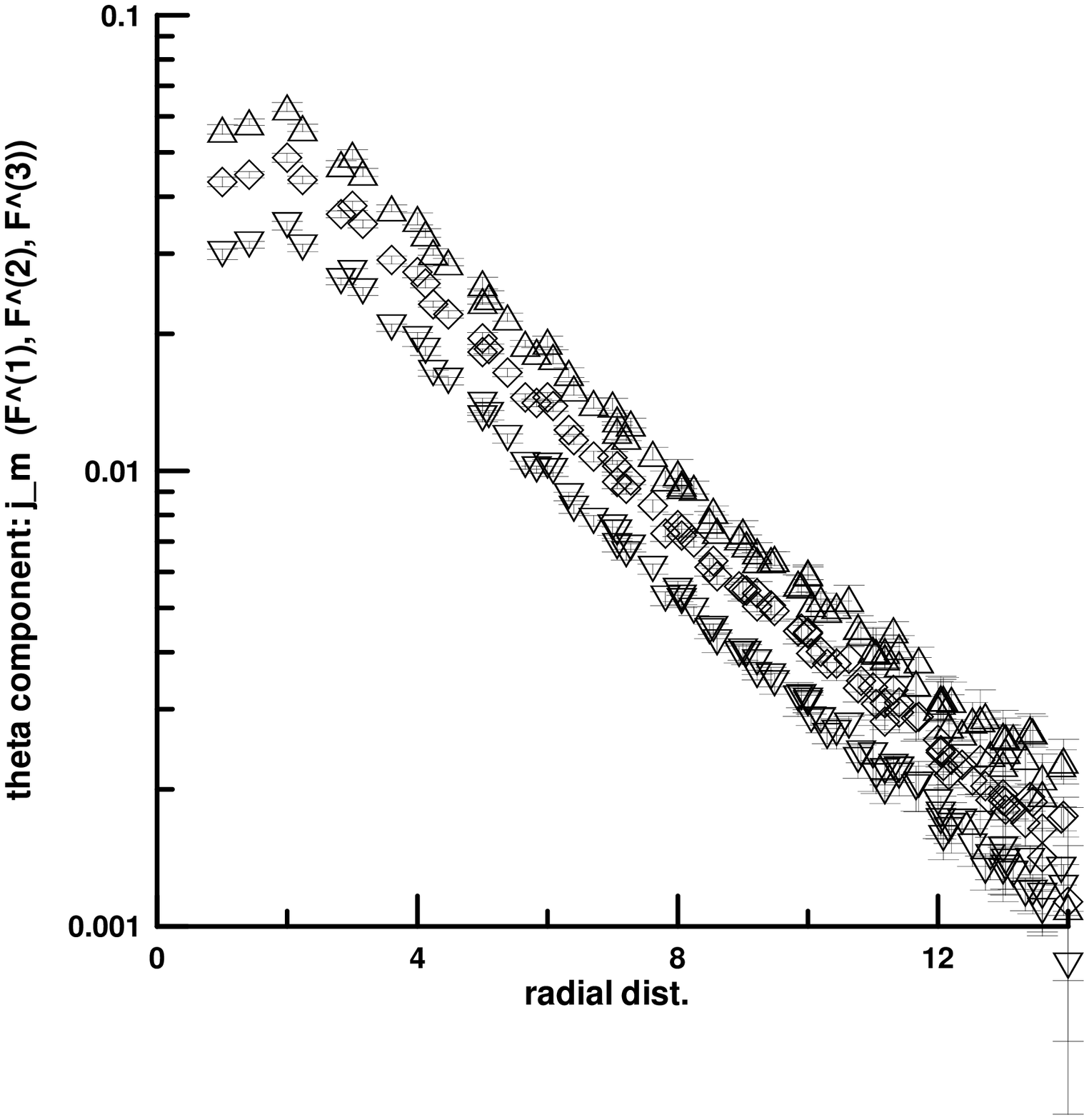}
Figure 2. Profile of the theta component of the magnetic current on the
mid-plane between $q$ and $\bar{q}$ separated by $13a$ based on (highest to
lowest) $\widehat{F}_{\mu\nu}^{(1)}$ (triangles), $\widehat{F}_{\mu\nu}^{(2)}$
(diamonds) and  $\widehat{F}_{\mu\nu}^{(3)}$ (inverted triangles).
\vskip 15mm

\end{document}